\documentclass[12pt]{article}
\usepackage{authblk}
\usepackage{blindtext}
\usepackage{graphicx} 

\usepackage{cite}
\usepackage{amsmath}
\usepackage{amssymb}
\usepackage{cleveref}
\usepackage{placeins}
\crefname{figure}{Fig.}{Fig.}
\Crefname{figure}{Fig.}{Fig.}
\usepackage{url}
\usepackage[legalpaper, portrait, margin=1in]{geometry}
\usepackage[width=0.88\textwidth]{caption}

\usepackage{nicefrac}       
\usepackage{microtype}      

\title{ 
A dual bistatic optical forward transceiver configuration for determining the position of an acoustic communication source detected by optical communication fibers.}
\author{Knut H. Grythe, Jan Erik Håkegård}
\affil{SINTEF Digital, Trondheim, Norway}
\affil{\text {\{knut.h.grythe, jan.e.hakegard\}@sintef.no}}

\begin{document}
\maketitle

\abstract{

Optical fibers have long been employed as sensors in a wide range of commercial systems, with applications such as distributed temperature and strain sensing. Distributed Acoustic Sensing (DAS) extends this concept by enabling the detection and localization of acoustic sources along the fiber, using backscattered light from small segments to achieve spatial resolution on the order of meters. Recently, DAS has been explored as a component in underwater acoustic communication systems, where the fiber conveys the acoustic field to a receiver at the fiber’s end. In addition to signal detection, DAS inherently provides source localization along the cable.
Emerging interest in bidirectional configurations—where both transmitter and receiver are placed at opposite ends of the fiber—has opened new possibilities. However, in such setups, source localization is not inherently integrated into the signal decoding process. For scenarios where source positioning is valuable, we propose a novel approach inspired by bi-static radar principles. This dual bi-static configuration utilizes acoustic signals received at both ends of the fiber to estimate source position based on propagation delay differences. Although the localization accuracy is lower than that of DAS due to reduced sampling rates, the method offers a viable alternative for integrated communication and positioning.
We present the system topology and configuration for a dual-fiber layout, each end equipped with separate optical transmitters and receivers. The position estimation is derived from the time difference of arrival (TDOA) between the two receivers. The Cramér-Rao Bound (CRB) is analytically derived to characterize the theoretical limits of localization accuracy, highlighting dependencies on system parameters such as optical power loss. Our analysis shows that increased acoustic bandwidth and higher carrier frequencies enhance spatial resolution.
Furthermore, we formulate the Cross Ambiguity Function (CAF) as a maximum likelihood estimator for TDOA and provide simulation results illustrating its performance under varying system conditions. Finally, we discuss key challenges that must be addressed for practical implementation, including the frequency-dependent conversion between the acoustic field and the optical signal.
}

\section{Introduction}
Optical fibers constitutes the backbone of modern data communication infrastructure, where long distance cables are connecting continents, while shorter units provide high speed local connections. At the same time, fibers are used as distributed sensors to detect different spatial phenomena like temperature, strain, vibration, and other parameters \cite{ding_advances_2023}. This capability finds it application in  various fields, including structural health monitoring like bridges, railway tracks, and industrial process monitoring. \cite{waagaard_real-time_2021}, \cite{wang_comprehensive_2019}. A specific application area is underwater acoustic monitoring where e.g. ships and other human generated noise sources are detected \cite{landro_sensing_2022}, in addition to marine animals like mammals \cite{rorstadbotnen_simultaneous_2023}, and earth quake \cite{shinohara_distributed_2025}. 

A much used technology for acoustic optical fiber sensing is the Rayleigh scattering based Distributed Acoustic Sensing (DAS) system, using an attached dedicated optoelectronic unit, normally called an interrogator \cite{IEEE-3103}, \cite{burdin_estimation_2022}. Such a system can detect and analyze external acoustic signals within fiber lengths in the order of 50 km depending upon the fiber signal losses and acoustic bandwidth, with a range resolution of typically a few meters. The latter is due to the very short optical pulses applied by the interrogator. 
A recent demonstrated application of the DAS is to detect and demodulate underwater acoustic communication signals \cite{potter_distributed_2024}, \cite{potter_first_2024}. This is an application spurred by the need for information flow from autonomous underwater vehicles (AUVs) to some control centers, either below or above sea surface. The excellent DAS range resolution is in this context important as it can be used to determine the position the AUV.
Similar to the back-scattering-based DAS technologies, conventional optical data communication systems can also be influenced by external acoustic vibrations —such as those caused by seismic activity — which induce micro-bending or strain in the fiber. These mechanical disturbances may result in signal attenuation, observable as amplitude variations at the receiver side. Such a forward sensing system based on optical communication is the topic of our work, where the acoustic source represents an underwater acoustic communication signal.     

While the DAS is able to estimate the position along the fiber of an external acoustic source, the communication receiver has no implicit position information available. given the received acoustic signal.

In the following, we explore the use of the received acoustic signal to estimate the position of the external acoustic source. This approach is motivated by two key factors. First, communication receivers are generally more cost-effective than dedicated interrogators. Secondly, the typical length of a DAS fiber is shorter than the distance between a pair of optical communication transceivers. However, as will be demonstrated, this method comes with a trade-off: the spatial resolution along the fiber of the estimated source position is lower. This is due to the limited acoustic bandwidth, which is order of magnitude smaller than the wide bandwidth available to DAS interrogator.

The basic principle of DAS is to detect the position of a source that is influencing the refraction of the fiber through causing reflection of a transmitted optical pulse. The position of the reflection point along the fiber is found through time-of-flight (ToF) principle, obtained through measuring time difference between emission of the pulse and its return to the transceiver, where the speed of light in a fiber is in the order of ${v_{f}} \approx 2*{10^8}\,\,m/s $. 

Communication is done via data modulating a transmitted light signal at the source end, and demodulating the data from the received light at the sink end. In the case where the propagation loss along the fiber is influenced by some external time variable acoustic field $\beta (t)$, the received optical signal can be formulated as $$r(t) = \beta (t)s(t) + {n_{receiver}}(t)$$ 
The signal amplitude variation due to $\beta (t)$ can be observed from $r(t)$. However, in contrast to the interrogator, there is no source position information embedded. 

The conversion between the external acoustic field and the observed amplitude variations at the receiver we assume is linear. This is a prerequisite to recreate the signal $\beta (t)$. The specific conversion mechanism and its properties, needs to be handled separately as discussed in section 6.    

In the following, various signaling tradeoffs, and future research topics are discussed addressing the concept and capabilities of using the acoustic signal for position estimation. In section 2 the basic principle of estimating the positioning is presented. The performance in terms of the Cramer-Rao bound (CRB) is deduced in section 3 for an acoustic communication signal. Section 4 illustrates the use of the cross ambiguity function (CAF) to find the position, while section 5 gives a set of numerical simulations. Section 6 presents some further research topics and challenges which need to be solved to obtain a harmonized and applicable system. The paper is concluded in section 7 with some statements about the findings .

\section{Communication fiber positioning principle}\label{sec:BasicConcept}
\begin{figure}[b!]
    \centering
    \includegraphics[width=1.0\textwidth]{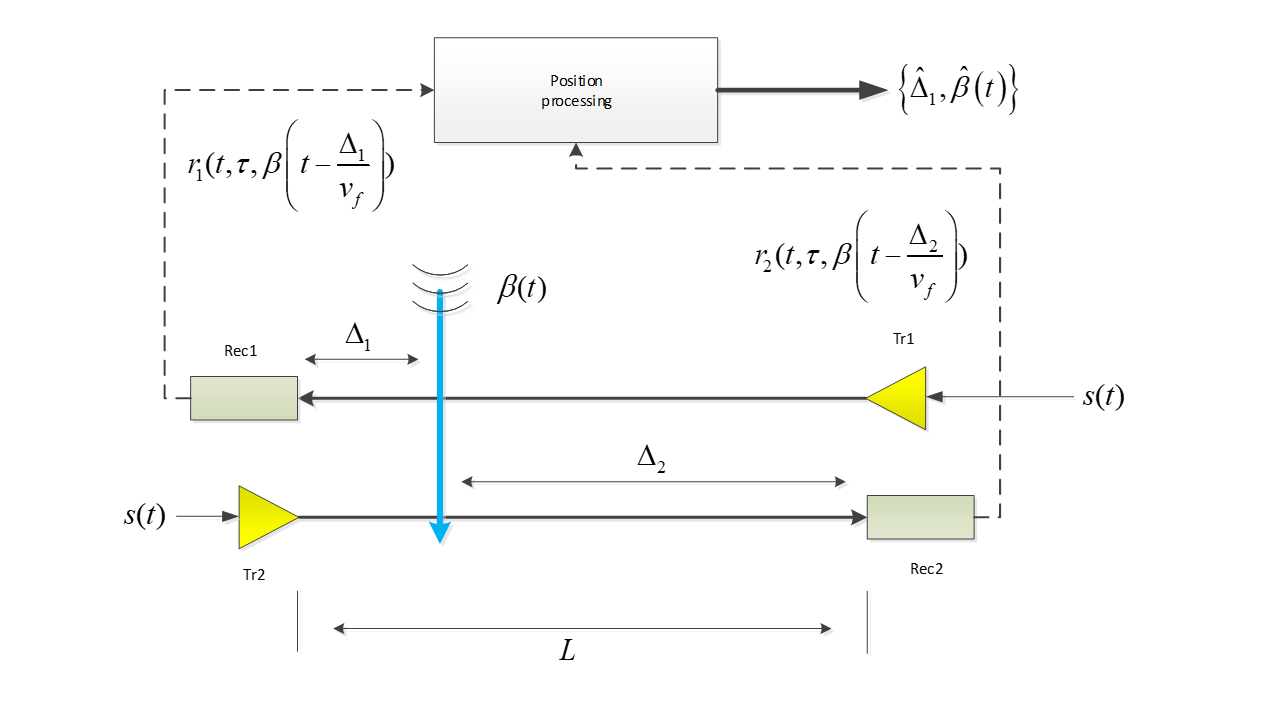}
    \caption{Overall optical - and analysis signaling flow}
    \label{fig:main_1}
\end{figure}

The estimation of the position of an external acoustic source sensed by a fiber equipped with optical communication transceivers, can conceptually be done from calculated time delays as illustrated in \cref{fig:main_1}. 
In \cref{fig:main_1} we have two fibers, where each fiber has a transmitter and receiver. The two fibers are bundled together and are configured to communicate in opposite directions where each end has a receiver - and transmitter pair. This can be considered to be dual bi-static configuration, using the terminology from radar literature. To jointly process the two detected acoustic amplitudes for position estimation, a third communication channel from one of the receivers to its transmitter end is necessary. This to convey its acoustic samples for jointly processing with the second set of acoustic samples, \cref{fig:main_1}. 

The position along the fiber relative to receiver 1, ${\Delta _1}$, of an acoustic source $\beta (t)$, can be expressed as follows applying the variables given in \cref{fig:main_1}. 

The time delays from the external source to the two receivers are $\tau_1 = \frac{\Delta_1}{v_f}$ and $\tau_2 = \frac{\Delta_2}{v_f}$. For positioning calculations, we are interested in the difference $\delta_\tau = \tau_1 - \tau_2$. Using these expressions, the distance to the acoustic source is calculated as
\begin{equation} \label{D1_eqn}
{\Delta _1} = \frac{{{\delta _\tau }{v_f} + L}}{2}
\end{equation}
The maximum and minimum values of the lengths and delays are 
\begin{equation} \label{D1max_eqn}
\Delta _1^{Max} = L\, \Rightarrow \delta _\tau ^{Max} = \frac{L}{{{v_f}}}
\end{equation}
and
\begin{equation} \label{D1min_eqn}
\Delta _1^{Min} = 0\, \Rightarrow \delta _\tau ^{Min} =  - \frac{L}{{{v_f}}}
\end{equation}

That is to determine the distance to the source, we must find the time difference ${\delta _\tau }$. 

The delay and position are found from a joint analysis of the two received signals 
\begin{equation} \label{r1t_eqn}
{r_1}\left(t,\tau ,\beta \left( {t - \frac{{{\Delta _1}}}{{{v_f}}}} \right)\right)
\end{equation}
and
\begin{equation} \label{r2t_eqn}
{r_2}\left(t,\tau ,\beta \left( {t - \frac{{{\Delta _2}}}{{{v_f}}}} \right)\right)
\end{equation}

To illustrate the analysis approach, we use a simplified signaling expression without any multipath component, and fiber self attenuation effects. Further the interaction between the acoustic signal and the fiber optical signal is linear in the amplitude. The signals from the transmitters are the same for both fibers, and equal to ${s\left( t \right)}$. The received signals are therefore expressed as

\begin{equation} \label{r1s_eqn}
{r_1}(t) = \beta (t - {\tau _1})s(t - {T_L}) + {n_{receiver}}(t)
\end{equation}
and
\begin{equation} \label{r2s_eqn}
{r_2}(t) = \beta (t - {\tau _2})s(t - {T_L}) + {n_{receiver}}(t)
\end{equation}
where ${T_L} = \frac{L}{{{v_f}}}$.

For $s(t)$ being a constant envelope/intensity signal like an optical carrier, the detection of $\beta (t)$ is done with an envelope detector giving the outputs
\begin{equation} \label{y1beta_eqn}
{z_1}(t,{\tau _1}) = \beta (t - {\tau _1}) + n_{_1}^\beta (t)
\end{equation}
and
\begin{equation} \label{y2beta_eqn}
{z_2}(t,{\tau _2}) = \beta (t - {\tau _2}) + n_2^\beta (t)
\end{equation}
  
The goal is to estimate the time difference ${\delta _\tau } = {\tau _1} - {\tau _2}$, from where the source position can be calculated. 

Based upon ${z_1}(t,{\tau _1})$ and ${z_2}(t,{\tau _2})$, an optimum value of ${\delta _\tau }$, ${\delta _\tau ^{Opt}}$, is found and the source position along the fiber is 
\begin{equation} \label{deltOpt_eqn}
\Delta _1^{Opt} = \frac{{\delta _\tau ^{Opt}{v_f} + L}}{2}
\end{equation}

\section{Positioning performance}

Prior to discussing the positioning estimator in the next section, we evaluate the relation between delay resolution and signaling bandwidth, and the lower bound on the quality of the delay estimator in terms of a Cramer Rao bound (CRB), where we assume that the estimator is unbiased.     

\subsection{Signaling bandwidth}
The resolution of the position deduction depends upon the speed of light in the fiber, ${v_{f}} \approx 2*{10^8}\,\,m/s $, and the bandwidth of $\beta (t)$, $B_\beta $. 
We assume that $B_\beta$ is limited to \[ - \frac{1}{{2{T_s}}} \leqslant B_\beta \leqslant \frac{1}{{2{T_s}}}\]
where $\beta (t)$ is equivalently base band sampled at the rate $\frac{1}{{{T_s}}}$. 
The spatial sampling resolution in meters along the fiber is then given as
\begin{equation}\label{equ_resol}
    {\Delta _\beta } = {v_f}{T_s}
\end{equation}
We find that for a sampling rate of ${10^4}\,Hz$, the spatial sampling resolution is 20 km, and 2 km for a sampling rate of ${10^5}\,Hz$. This indicates that to determine the source position with a resolution lower than e.g. 1 km, a sufficiently high signal bandwidth and a good estimator are required.

\subsection{Cramer-Rao bound}
As given in \eqref{deltOpt_eqn}, the position is determined using the estimation of ${\delta _\tau }$. 
The quality of an unbiased parameter estimator is lower bounded by the Cramer-Rao bound (CRB) \cite{harry_l_van_trees_detection_1968-1}. To find the CRB for the ${\delta _\tau }$ estimator, we apply the signal representation and CRB deduced results from \cite{delmas_cramer_2012}.

The received acoustic communication envelope signal is a bandpass signal  \[\beta \left( t \right) = \operatorname{Re} \left( {{s^{Ac}}\left( t \right){e^{j2\pi {f_0}t}}} \right)\]  centered around a carrier frequency ${f_0}$. Including the bandpass noise signals, the two outputs are expressed as   
\[{z_1}\left( t \right) = \beta \left( t \right) + n_{_1}^\beta \left( t \right)\]
\[{z_2}\left( t \right) = a\beta \left( {t - {\tau _0}} \right) + n_{_2}^\beta \left( t \right)\]
where $a$ and ${{\tau _0}}$ denote the relative attenuation and delay of $\beta \left( t \right)$ at receiver 2 with respect to receiver 1 in \cref{fig:main_1}.
Actually in our case, $\beta (t)$ should have been written as $\beta (t\left| {{\Delta _1},{f_0}} \right.)$ since we have a distance dependency in the expected received signal power as
\begin{equation}\label{equ_fiber_gain_sens}
    E{\left| {\beta (t)} \right|^2} = g\left( {{\Delta _1},{f_0}} \right)
\end{equation}
This is due to the  power loss factor of the fiber optical cable in terms of $dB/km$, and the frequency dependent sensing of the fiber. We omit expressing this conditioned dependency explicitly in the theoretical framework.  

After bandpass filtering and down conversion to baseband, two complex envelopes are obtained as 
\[{u_1}\left( t \right) = {s^{Ac}}\left( {t - {\tau _g}({f_0})} \right) + {n_1}\left( t \right)\]
\[{u_2}\left( t \right) = \kappa {s^{Ac}}\left( {t - {\tau _g}({f_0}) - {\tau _0}} \right) + {n_2}\left( t \right)\]

where $\kappa  = a{e^{ - j2\pi {f_0}\left( {{\tau _0} + {\tau _\phi }({f_0})} \right)}}$. Here ${{\tau _g}({f_0})}$ and ${{\tau _\phi }({f_0})}$ represent the group delay and phase delay of the bandpass filter, respectively.
The complex baseband signals ${{s^{Ac}}\left( t \right)}$, ${n_1}\left( t \right)$ and ${n_2}\left( t \right)$ are assumed to be circular, mutually uncorrelated, zero mean Gaussian and wide sense stationary (WSS). 

The CRB calculations use the corresponding acoustic power spectral densities (PSD) given as $S\left( f \right)\,$,  ${N_1}(f)$,  and ${N_2}(f)$.  

Based upon these signal representations, the delay CRB is given as \cite{delmas_cramer_2012}

\begin{equation}\label{equ_delay_crb}
CRB\left( {{\tau _0}\left| {N,\kappa } \right.} \right) = {\mathbf{J}}_{{\tau _0},{\tau _0}}^{ - 1} = {\left( {8{\pi ^2}{{\left| \kappa  \right|}^2}N\int_{ - \frac{1}{{2{T_s}}}}^{\frac{1}{{2{T_s}}}} {\frac{{{f^2}{S^2}(f)}}{{{\Delta _\kappa }(f)}}df} } \right)^{ - 1}}
\end{equation}

where $\left| \kappa  \right|=a$ and ${\Delta _\kappa }(f) = \left( {S(f) + {N_1}(f)} \right)\left( {S(f) + {N_2}(f)} \right) - {\left| \kappa  \right|^2}{S^2}\left( f \right)$. $N$ is the number of consecutive time samples of ${u_1}\left( {k{T_s}} \right)$ and ${u_2}\left( {k{T_s}} \right)$. 
  
The term ${{\bf{J}}_{{\tau _0},{\tau _0}}}$ represents a diagonal factor in the Fischer Information Matrix (FIM), \cite{harry_l_van_trees_detection_1968-1}. 

The bound is not a function of ${{\tau _0}}$, but for our case the dependency is implicit through the fiber range loss factor, given by the variation of $\left| \kappa  \right|$. Further we observe that the number of samples $N$ can be used to balance the variations in $\left| \kappa  \right|$. 

\subsubsection{Numerical illustrations}
To visualize the behavior of the CRB versus $\left| \kappa  \right|$ and $N$, we have made a two simple simulations where the PSDs are assumed to be constant, and setting ${T_s} = 1$.  
Carrying out the integration, the CRB for this case is 
\begin{equation}\label{equ_flat_crb}
CR{B^{Flat}}\left( {{\tau _0}\left| {{T_s}} \right.} \right) = {\left( {\frac{2}{3}\frac{1}{{T_s^3}}{\pi ^2}{{\left| \kappa  \right|}^2}N\frac{{{S^2}}}{{\left( {1 - {{\left| \kappa  \right|}^2}} \right){S^2} + {N_1}{N_2} + S{N_1} + S{N_2}}}} \right)^{ - 1}}
\end{equation}

Since $CRB\left( {{\tau _0}} \right) \ge 0$ we require 
\begin{equation}\label{equ_kappa_value}
\left( {1 - {{\left| \kappa  \right|}^2}} \right) \ge 0 \Rightarrow 0 \le \left| \kappa  \right| \le 1
\end{equation}
This requirement makes $CRB\left( {{\tau _0}} \right) \ge 0$ independent of the levels of the PSDs.   

The numerical values applied for the two simulations are the cable length equal to $60$ km; the noise level of $0$ dB is given at the receivers output; the signal level of $10$ dB given is defined to be at the receiver 1 output when the  $\beta \left( t \right)$ position is at the receiver 1 input. This means that the received signal level or SNR are scaled with the $\beta \left( t \right)$ position along the fiber due to the fiber signal losses.

\begin{figure}[t!]
    \centering
    \includegraphics[width=0.7\textwidth]{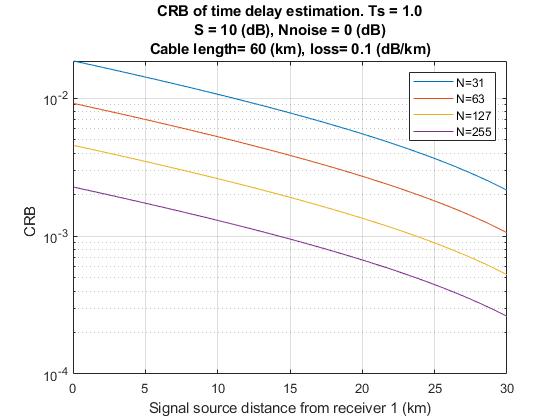}
    \caption{CRB for different PN sequence lengths, with a fiber loss of 0.1 dB/km}
    \label{fig:crb_01_loss}
\end{figure}

\begin{figure}[t!]
    \centering
    \includegraphics[width=0.7\textwidth]{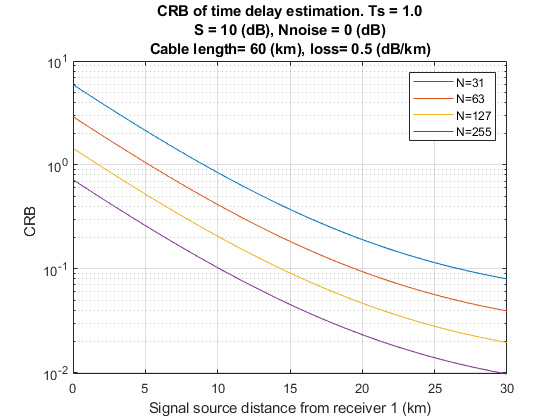}
    \caption{CRB for different PN sequence lengths, with a fiber loss of 0.5 dB/km}
    \label{fig:crb_05_loss}
\end{figure}

The CRB results are presented in \cref{fig:crb_01_loss} and \cref{fig:crb_05_loss} where the fiber losses are set to 0.1 and 0.5 $dB/km$, respectively. The number of samples $N$ are 31, 63, 127 and 255. This corresponds to a set of possible pseudo noise (PN) sequence lengths.

From \cref{fig:crb_01_loss} and \cref{fig:crb_05_loss}, three important observations can be deduced.

\begin{figure}[t!]
    \centering
    \includegraphics[width=0.7\textwidth]{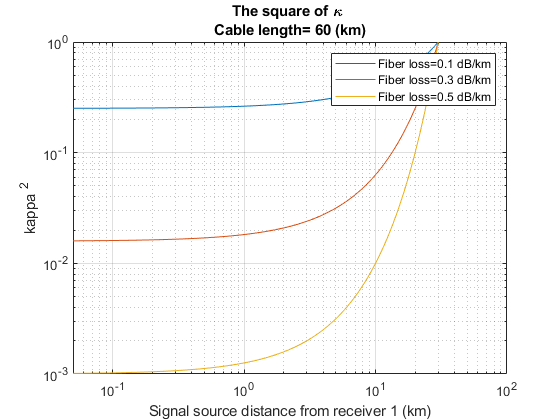}
    \caption{The square of $\kappa $ for three different fiber loss factors}
    \label{fig:kappa2}
\end{figure}

\begin{itemize}
    \item The CRB is decreasing to a minimum at the center of the cable
\end{itemize}
\begin{itemize}
    \item The CRB is decreasing with higher values of $N$
\end{itemize}
\begin{itemize}
    \item The CRB is increasing with higher cable loss factors. 
\end{itemize}
The CRB is strongly influenced by the fiber loss factor. The loss factor is causing ${{{\left| \kappa  \right|}^2}}$ to vary with the distance from the center of the cable, where it is equal to 1.0. \cref{fig:kappa2} illustrates this showing ${{{\left| \kappa  \right|}^2}}$ for three different values of the fiber loss factors. 

We find that large values of the loss factor limits the applicability of the position estimation to the center of the fiber. CRB values larger than $1.0$ are not applicable. In general, CRB values much smaller than the Nyquist sampling period are usually required or wanted. For a practical system design, an upper bounded CRB, $CR{B_{UpperBound}}\left( {{\tau _0}} \right)$, is normally specified. This leads to a parameter subset defined as  
\begin{equation}\label{equ_crb_bound}
{\rm A} \equiv \left\{ {\left| {{\kappa _b}} \right|,{N_b}} \right\} \Rightarrow CRB\left( {{\tau _0}\left| {{\rm A},S,\,{N_1},\,{N_2}} \right.} \right) \le CR{B_{UpperBound}}\left( {{\tau _0}} \right)
\end{equation}
That is, for a given $CR{B_{UpperBound}}\left( {{\tau _0}} \right)$ and signal parameters $S$, ${{N_1}}$, and ${{N_1}}$, the subset ${\rm A} \equiv \left\{ {\left| {{\kappa _b}} \right|,{N_b}} \right\}$ represents the possible valid space for the variables $\left| \kappa  \right|$ and $N$.

\section{Delay estimation} 
\subsection{Cross ambiguity function}\label{sec:ambfun}
The acoustic source will normally be carried by e.g. an underwater autonomous vehicle (AUV). The observed acoustic source may therefore be expected to be oriented at a random position along the cable. Consequently, the apriori time difference probability density function, $p(\tau )$, is uniform 
\begin{equation}\label{equ_pdf_tau}
p(\tau ) = \frac{{{v_f}}}{{2L}}\tau \,\,\,,\, - \frac{L}{{{v_f}}} \leqslant \tau  \leqslant \frac{L}{{{v_f}}}
\end{equation}
This causes the optimum estimator of the position to be a maximum likelihood estimator (MLE), \cite{harry_l_van_trees_detection_1968-1}. The estimation of the time difference is done applying a cross correlation analysis between the two signals received at each cable end.
Generally, the output of the cross correlation is also used for detecting the presence of the communication signal via the binary receiver operating characteristics (ROC) statistics \cite{harry_l_van_trees_detection_1968-1}, prior to time estimation. However, since we in this paper are not estimating the source position using the CAF, the ROC behavior is not analyzed, but left for future system work.

From the previous CRB analyses, we find that the CRB performance is determined by three parameters; signal SNR at receiver 1, power ratio between receiver 1 and receiver 2 outputs, and the number of samples applied in the estimator. To illustrate the dependence of the cross correlation on these three parameters, we calculate the cross ambiguity function (CAF) of an acoustic band limited signal, oriented at different positions along the fiber between receiver 1 and the cable mid point. 
  
We assume that the optical receivers are capable of linearly reconstructing the acoustic baseband signal ${s^{Ac}}(t)$. The calculation of the CAF is done on the receivers acoustic matched filter output $$y(t) = {s^{Ac}}(t) * h_R^{MF}(t) + n(t)$$ or expressed as $y(t) = x(t) + n(t)$. The CAF is generally written as   
\begin{equation}\label{equ_gen_ambig}
\left| {{\Psi _{cross}}\left( {\tau ,\nu } \right)} \right| \buildrel \Delta \over = \left| {\int\limits_{ - \infty }^\infty  {{y_1}\left( t \right)y_2^ * \left( {t + \tau } \right){e^{j2\pi \nu t}}dt} } \right|
\end{equation}
where $\tau $ is the delay offset, in our case equal to ${\delta _\tau } = {\tau _1} - {\tau _2}$ , and $\nu $ is a possible Doppler shift between receiver 1 and receiver 2.
With respect to the applicability of the CAF, the CAF represents the maximum likelihood estimator (MLE) function of the delay and Doppler for the case of slowly fluctuating signals or radar target \cite{harry_l_van_trees_detection_1972}. 

The optimum values of ${\delta _\tau } = {\tau _1} - {\tau _2}$ and $\nu $  are found as

\begin{equation} \label{opt_caf_eqn}
\left\{ {\delta _\tau ^{Opt},\,{\nu ^{Opt}}} \right\} \equiv \mathop {\max }\limits_{\delta _\tau ^{Min} \le {\delta _\tau } \le \delta _\tau ^{Max};\,\nu  \in D(\nu )} \left( {\left| {{\Psi _{cross}}\left( {\tau ,\nu } \right)} \right|} \right)
\end{equation}

where $\delta _\tau ^{Max}$ and $\delta _\tau ^{Min}$ are defined by equ. (\ref{D1max_eqn}) and equ. (\ref{D1min_eqn}). $D\left( \nu  \right)$ represents the domain of $\nu $.  

\subsection{CAF terms}

When the CAF is decomposed in its individual terms, there is four terms. These are $$\left\{ {{x_1}(t)x_2^ * \left( {t + \tau } \right);\,\,{n_1}(t)x_2^ * \left( {t + \tau } \right);\,\,{x_1}(t){n_2}(t);\,\,{n_1}(t){n_2}(t)} \right\}$$
There is one signal term only, and three terms involving the additive noise. The signal term represents scaled and delayed versions the same acoustic signal, providing the autocorrelation. 

For the expectations and variances of the terms involving noise, all the factors are statistically independent. Therefore we get for the noise only term

\begin{equation}\label{equ_e_noise}
E\left( {{n_1}(t){n_2}(t)} \right) = E\left( {{n_1}(t)} \right)E\left( {{n_2}(t)} \right) = {\mu _{n1}}{\mu _{n2}}
\end{equation}

\begin{equation}\label{equ_var_noise}
Var\left( {{n_1}(t){n_2}(t)} \right) = \left( {\sigma _{n1}^2 + \mu _{n1}^2} \right)\left( {\sigma _{n2}^2 + \mu _{n2}^2} \right) - \mu _{n1}^2\mu _{n2}^2
\end{equation}
The pdf of this noise product term involves the modified Bessel function of the second kind. \cite{gaunt_basic_2022}.    

The acoustic signal $x(t)$ we assume has a mean value of zero, with the results

\begin{equation}\label{equ_x1_mean}
E\left( {\,{n_1}(t)x_2^ * \left( {t + \tau } \right)} \right) = 0
\end{equation}

\begin{equation}\label{equ_x1_var}
Var\left( {{n_1}(t)x_2^ * \left( {t + \tau } \right)} \right) = \left( {\sigma _{n1}^2 + \mu _{n1}^2} \right)\sigma _{x2}^2
\end{equation}

\begin{equation}\label{equ_x2_mean}
E\left( {{x_1}(t){n_2}(t)} \right) = 0
\end{equation}

\begin{equation}\label{equ_x2_var}
Var\left( {{x_1}(t){n_2}(t)} \right) = \left( {\sigma _{n2}^2 + \mu _{n2}^2} \right)\sigma _{x1}^2
\end{equation}

The variance of each of the four terms varies depending on the position of the acoustic source along the fiber cable.

The autocorrelation term of $E\left( {{x_1}(t)x_2^ * \left( {t + \tau } \right)} \right)$ is amplitude scaled and delay shifted depending upon the position along the cable. In the case where the signal is a PN sequence, the size of the correlation peak value varies with the position.  
The noise variance of (\ref{equ_var_noise}) is assumed independent of the source position since it represents the white noise at the output of each receiver. 
The two combined noise and signal terms of (\ref{equ_x1_var}) and (\ref{equ_x2_var}) are scaled with the cable loss in a counter acting way, as can be seen from \cref{fig:kappa2}.

We find that the CAF contains a fixed variance noise term, and three other terms depending on the source position.

\subsection{Time discrete CAF}
To express the CAF in terms of samples, a time discrete version is applied. 
\begin{equation}\label{equ_disc_ambig}
\left| {\Psi _{cross}^{Ts}\left( {\tau \left| \nu  \right.} \right)} \right| \triangleq \left| {\frac{1}{K}\sum\limits_{k = 0}^{K - 1} {\left( {({x_1}(k) + {n_1}(k))(x_2^ * \left( {k + \tau } \right) + {n_1}(k)){e^{j2\pi \nu k}}} \right)} } \right|
\end{equation}
where $\nu $ is the given Doppler, and the number of sample $K$ is assumed to be enough to encapsulate the entire sequence set $\{ x(k)\} $.  This discrete CAF is normalized with the number of samples to illustrate the impact of $K$ on the CAF fluctuations. 

Consider the estimate of the mean value of a set of time samples $\{ b(m)\}$ is given as  \[{{\hat \mu }_b} = \frac{1}{M}\sum\limits_{m = 0}^{M - 1} {b(m)} \]. 

Given the variance of $\{ b(m)\} $ as $\sigma _b^2$, and assuming ${\mu _b} = 0$, the variance of ${{\hat \mu }_b}$ is equal to \[Var[{{\hat \mu }_b}] = \frac{1}{M}\sigma _b^2\]
From this we expect that the overall fluctuation of the CAF decrease as the number of samples, $K$, increases.

\section{Numerical CAF performance}
To illustrate the behavior of the CAF under varying conditions of the primary parameters which impact the CRB, we present simulation results through a series of graphical representations. The representations are the linear CAF values, along with the complementary cumulative distribution function (CCDF) of the CAF values within the area in the $\{ \tau ,\,\nu \} $ domain surrounding the main peak. The CCDF indicates to what extent the CAF amplitude elements surrounding the peak element exceed a threshold. This is heavily influencing the ROC performance, and hence the ability to obtain a reliable delay estimate. 
According to \eqref{opt_caf_eqn}, the optimum estimate of delay and Doppler is found from the maximum of the CAF. Large parasitic peaks surrounding the true maximum, may mislead the ROC hypothesis testing \cite{harry_l_van_trees_detection_1972}{}, and give a wrong and biased delay estimate. Thus, a high probability of such peaks is reducing the probability of a near CRB estimation value.   

\subsection{Simulated system and values}
The CAF simulations are carried out on a fiber cable with a total length of 60 km, with 30 km to the midpoint.  
The fiber power loss set are limited to the range of 0.05 to 0.5 dB/km, respectively, where the sub sets are linked to the specific simulation.

The acoustic signal sensed by the cable, $\beta \left( t \right)$, is chosen to be a band pass BPSK modulated data signal on a given carrier frequency, ${f_0}$. The data consists of pseudo noise (PN) sequences, with sequence lengths ranging from 31 to 511 chips, used to illustrate the number of samples of the CRB. The sampling rate is 16 KSps with an oversampling of 4, and a cosine roll off filter with roll-off equal to 0.31 and 8 symbols long.  

The SNR values provided are defined as those observed at the output of receiver 1 for an acoustic source positioned at the same end of the fiber. Consequently, the SNR value observed at receiver 1 will decrease as the acoustic source position moves from receiver 1 towards the midpoint of the cable. Once the source passes the midpoint, receiver 2 becomes the reference receiver since the receiver 2 then will observe a stronger signal. 

The variations of the given SNR value for a given position can be interpreted in two ways. Either as a variation in the power of the acoustic source it self, or due to a frequency dependent coupling efficiency from the acoustic to the optical domain. A reduced efficiency at higher frequencies, will lower the available SNR. The SNR value is a functional one dimensional mapping from this two-dimensional set. 

\subsection{Reference CAF signaling performance }
In the following two sub sections, the performance of the CAF is illustrated for different fiber cable losses, and PN sequence lengths, respectively. To provide a set reference performance curves, we have in the present sub section simulated the CAF, and its CCDF performance under ideal two noise conditions, for a fixed PN length. That is; for no noise, and noise only ($SNR =  - 70\,\,dB$). The PN length is equal to 127 chips. The position of the source is set to be near receiver 1 since these two modes of operation is independent of the source position.  

\begin{figure}[t!]
    \centering
    \includegraphics[width=0.5\linewidth]{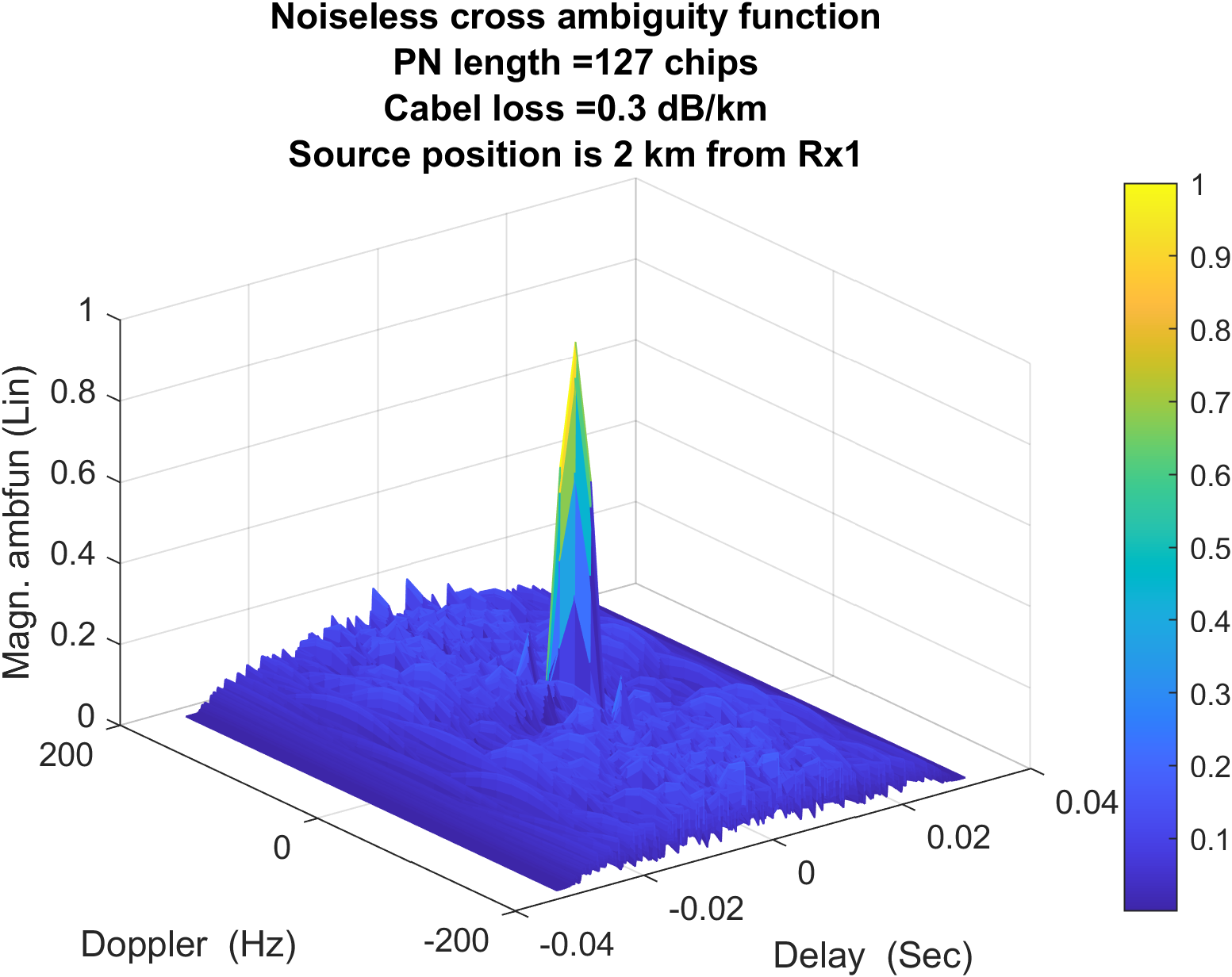}
    \caption{Noiseless CAF with PN length = 127 chips}
    \label{fig:ref_noiseL_CAF}
\end{figure}

\begin{figure}[t!]
    \centering
    \includegraphics[width=0.5\linewidth]{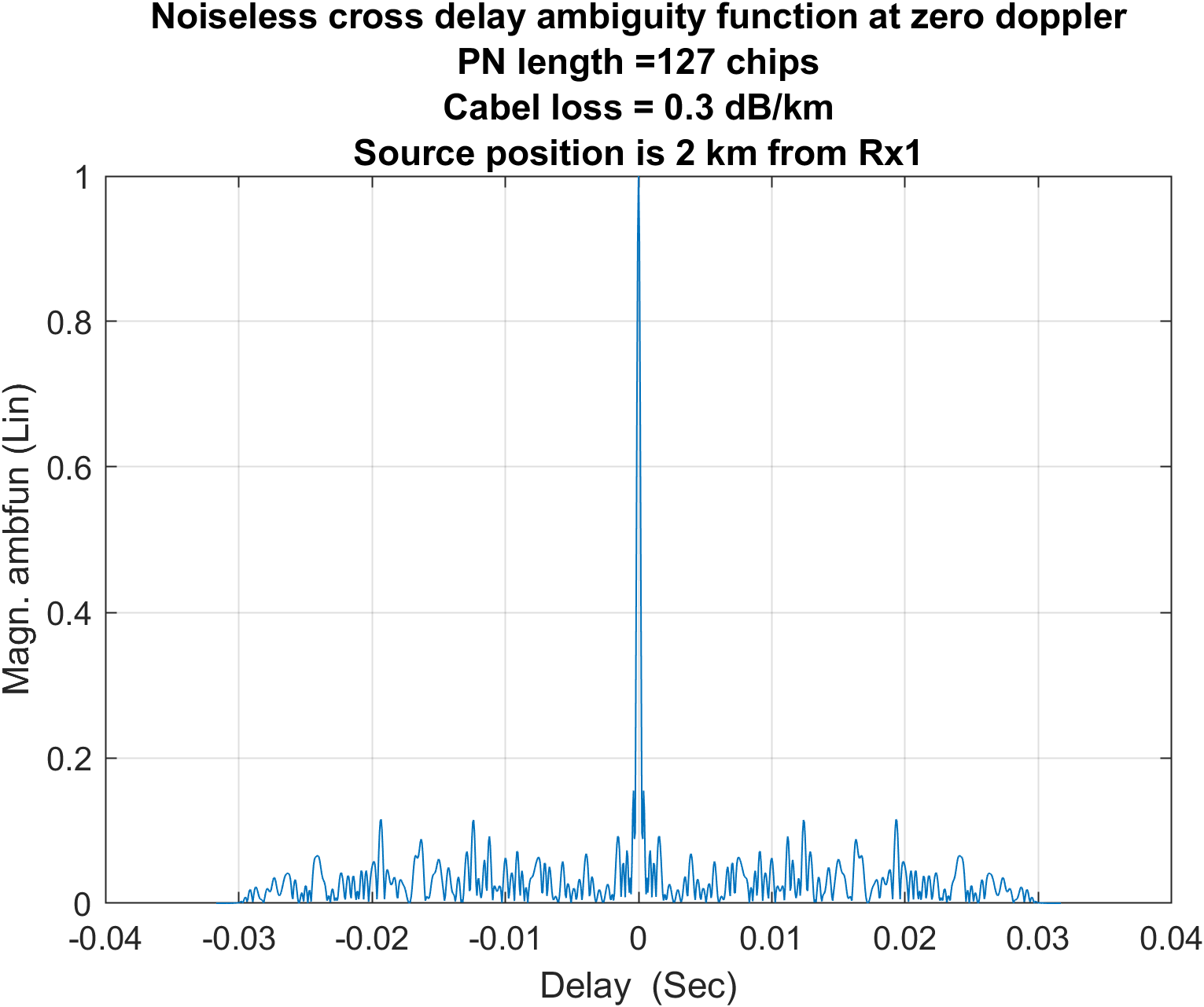}
    \caption{Noise less delay of CAF at zero Doppler}
    \label{fig:ref_noiseL_delay}
\end{figure}

\begin{figure}[t!]
    \centering
    \includegraphics[width=0.5\linewidth]{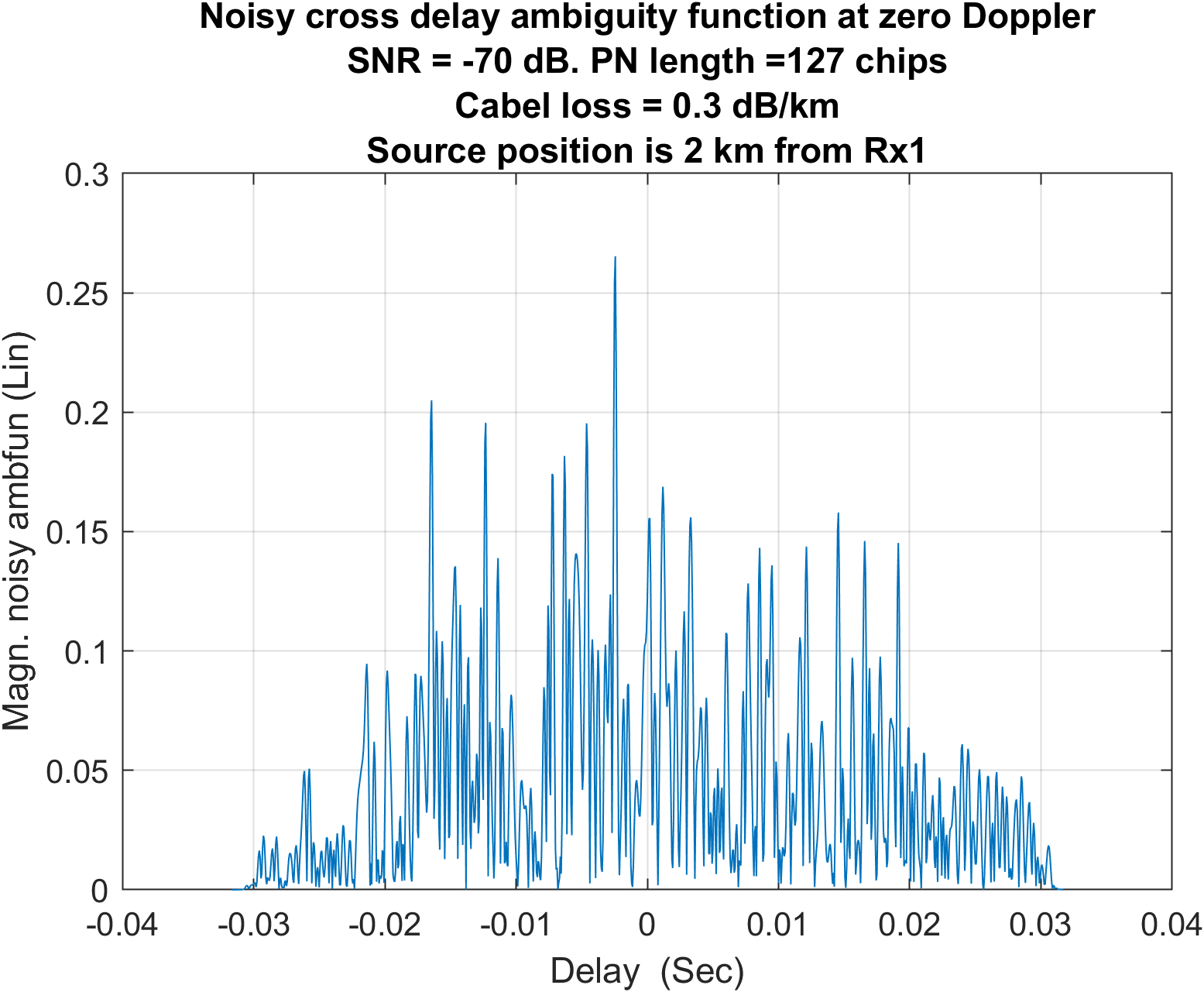}
    \caption{Noisy delay of CAF at zero Doppler}
    \label{fig:ref_noisy_delay}
\end{figure}

\begin{figure}[b!]
    \centering
    \includegraphics[width=0.5\linewidth]{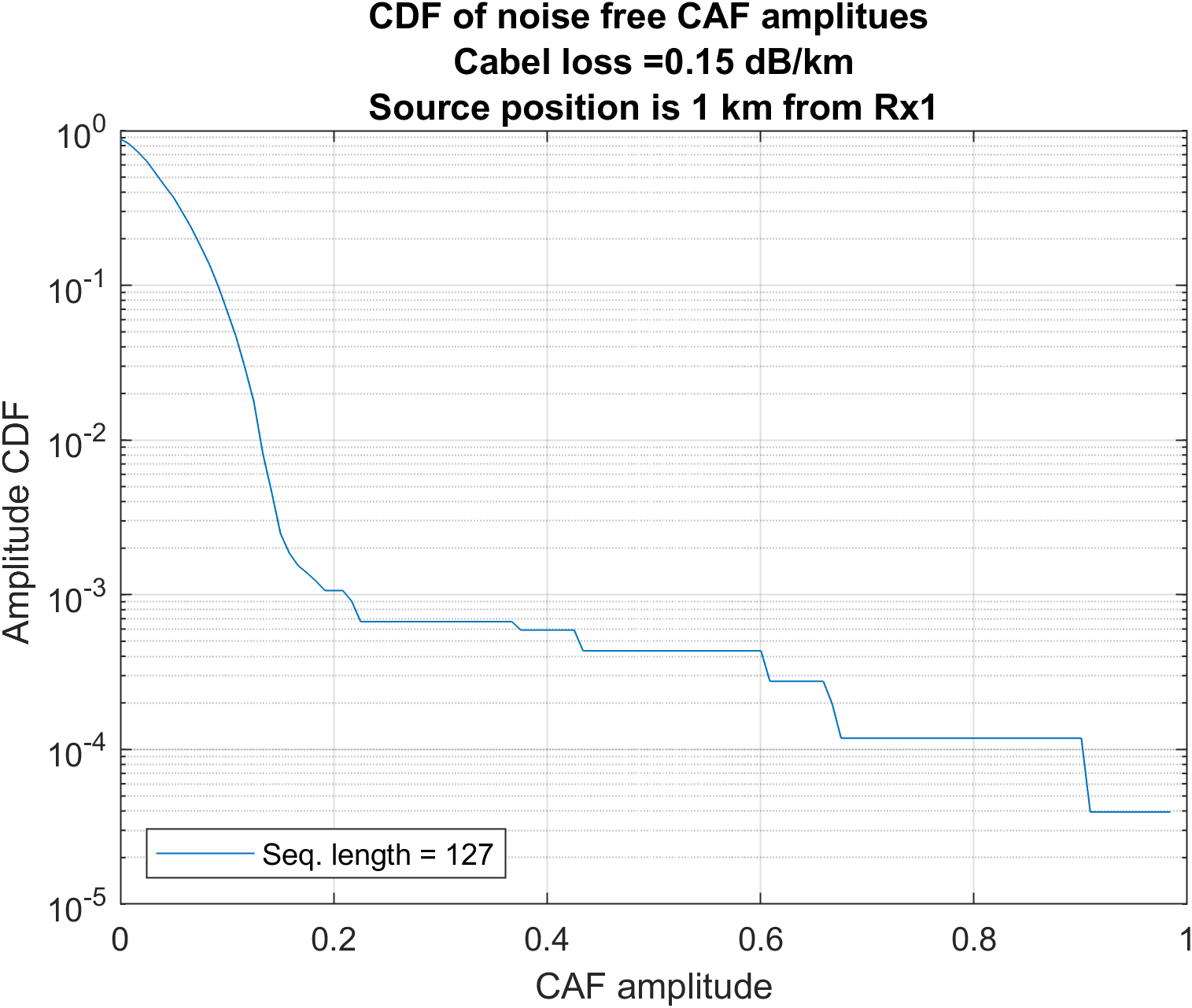}
    \caption{Noise free CCDF of CAF amplitudes}
    \label{fig:ref_noiseL_CCDF}
\end{figure}

\begin{figure}[t!]
    \centering
    \includegraphics[width=0.5\linewidth]{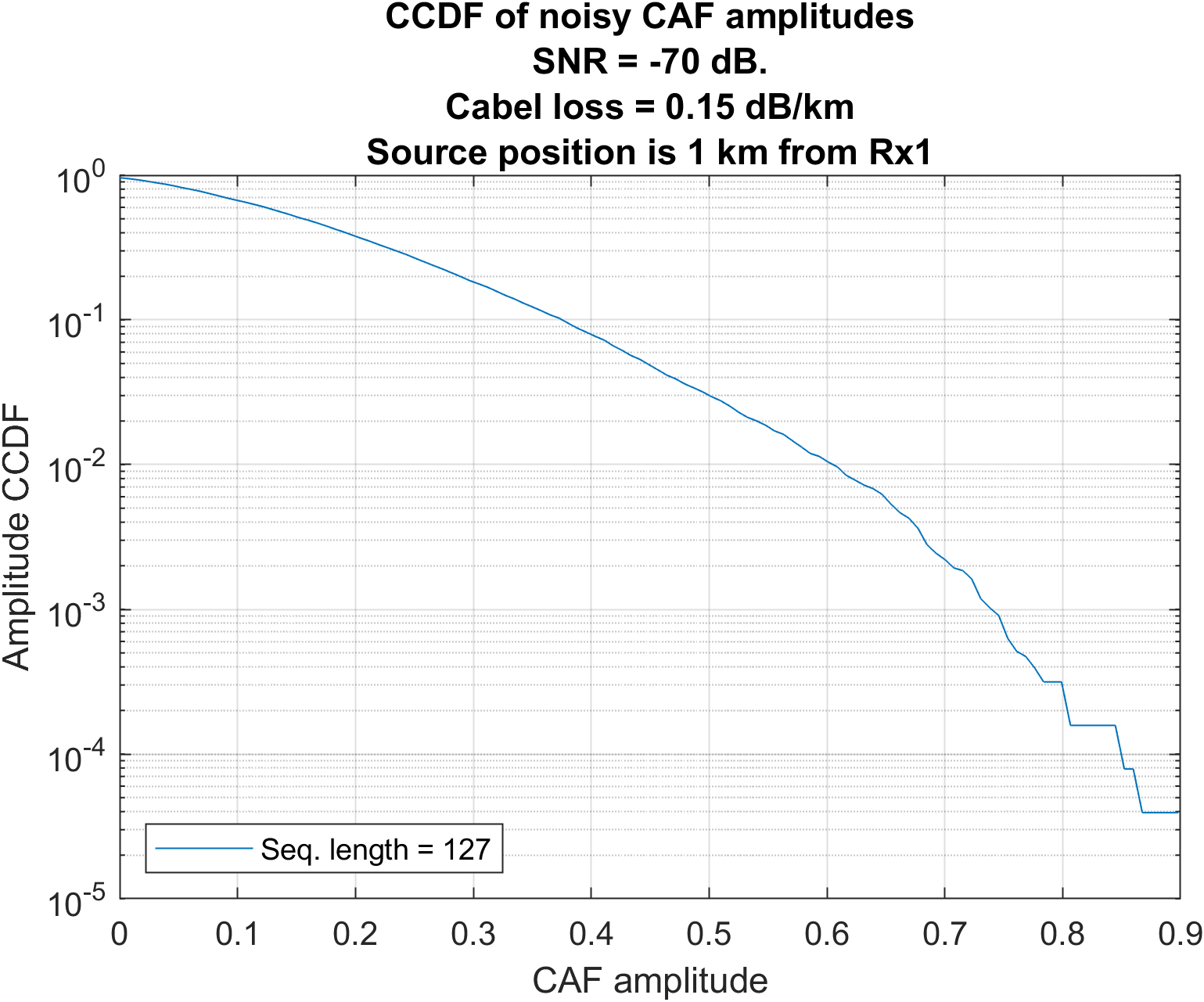}
    \caption{Noise only CCDF of CAF amplitudes}
    \label{fig:ref_noisy_CCDF}
\end{figure}

The five figures, \cref{fig:ref_noiseL_CAF} to \cref{fig:ref_noisy_CCDF}, clearly illustrate the shift of the CAF properties as the conditions go from 'no noise' to 'noise only' at the receivers output; from a single main peak in the Delay/Doppler domain to a full set of parasitic peaks. 

\subsection{Varying fiber losses with noise}

One of the important parameters in the delay CRB expression, is $\left| \kappa  \right|$ or equivalently the fiber loss factor in dB/km, equ. (\ref{equ_delay_crb}). We therefore include a set of CAF visualizations, and CCDF simulations where the corresponding cable loss is varying from 0.05 to 0.5 dB. The PN sequence is allocated a length of 127 chips for all simulations. The source positions are 29 km and 1 km, respectively. The reference SNR is set to 10 dB.


\begin{figure}[b!]
    \centering
    \includegraphics[width=0.5\linewidth]{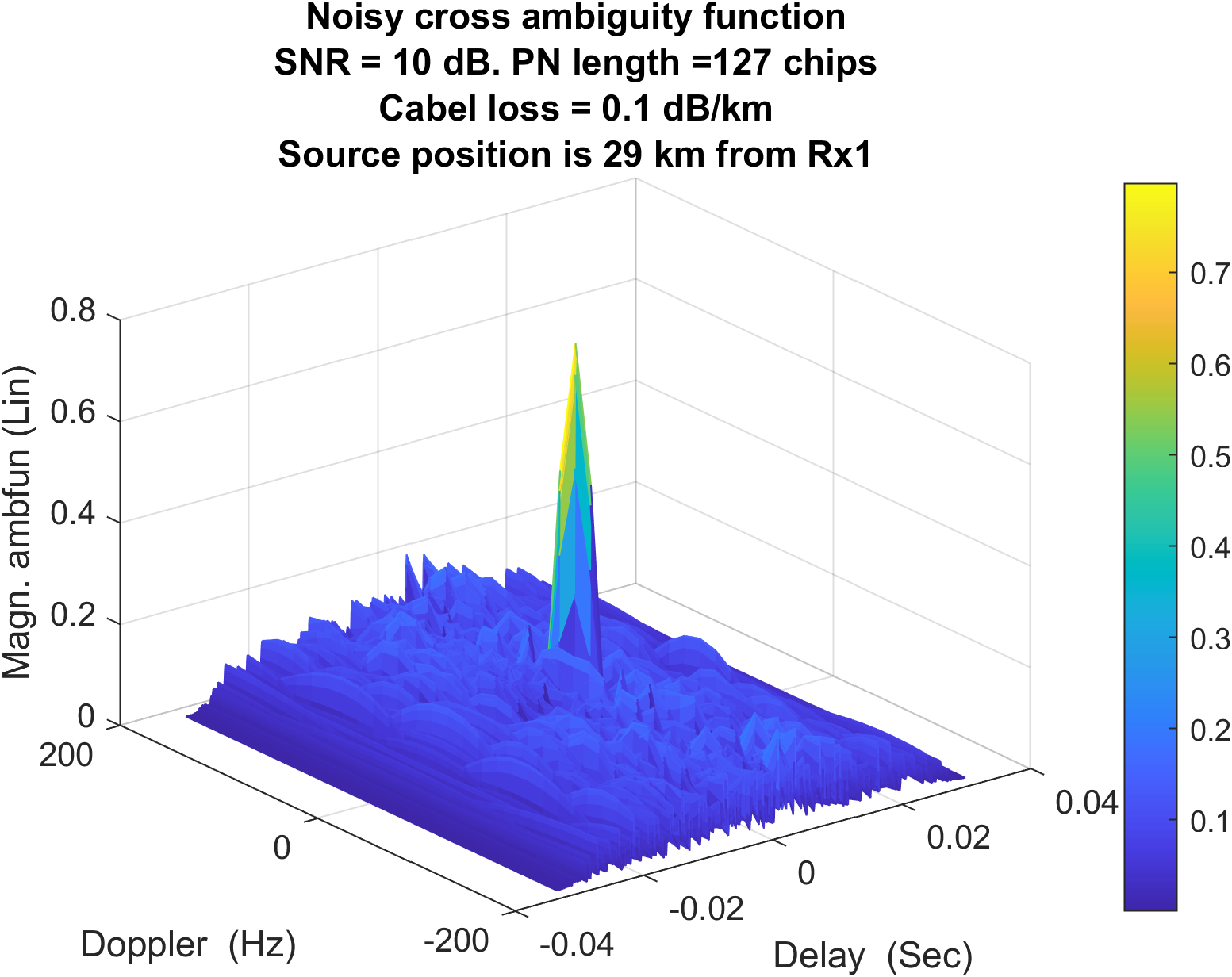}
    \caption{CAF at position 29 km. Cable loss equal to 0.1 dB/km}
    \label{fig:fibL_CAF_middle_01}
\end{figure}


\begin{figure}[ht!]
    \centering
    \includegraphics[width=0.5\linewidth]{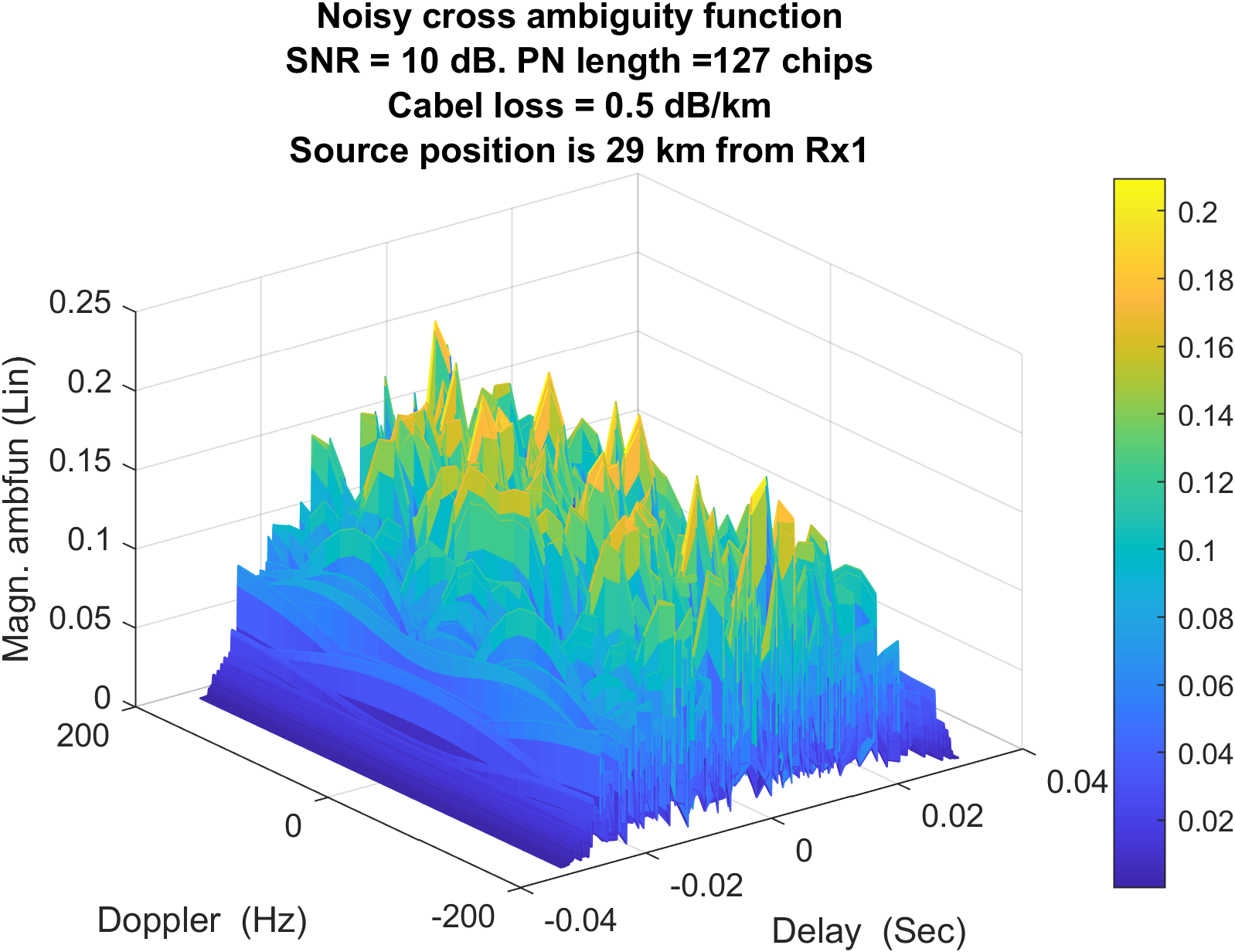}
    \caption{CAF at position 29 km. Cable loss equal to 0.5 dB/km}
    \label{fig:fibL_CAF_middle_05}
\end{figure}


\begin{figure}[ht!]
    \centering
    \includegraphics[width=0.5\linewidth]{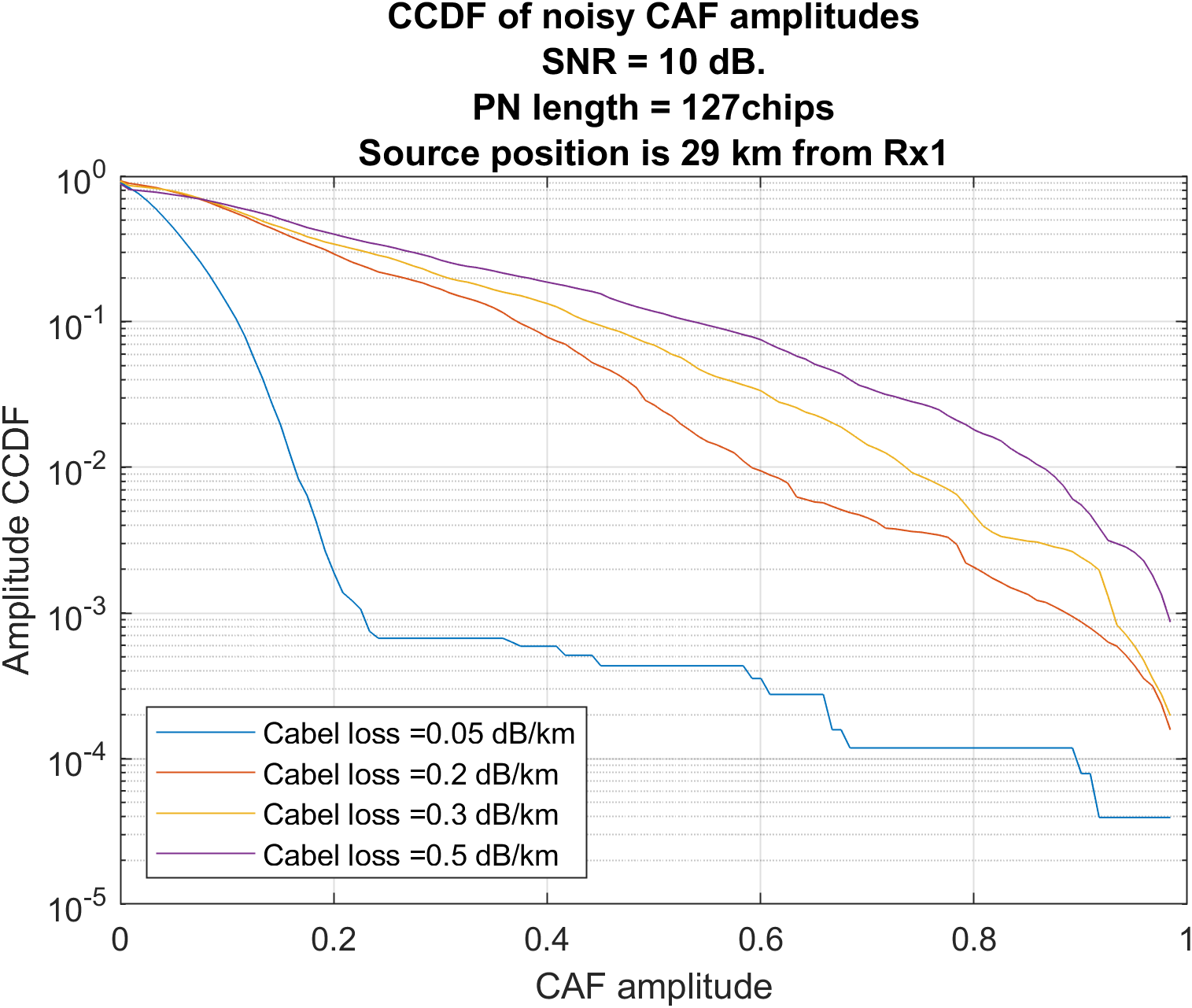}
    \caption{CAF amplitude CCDF at position 29 km for a PN length equal to 127 chips}
    \label{fig:fibL_CCDF_middle_127}
\end{figure}

\begin{figure}[ht!]
    \centering
    \includegraphics[width=0.5\linewidth]{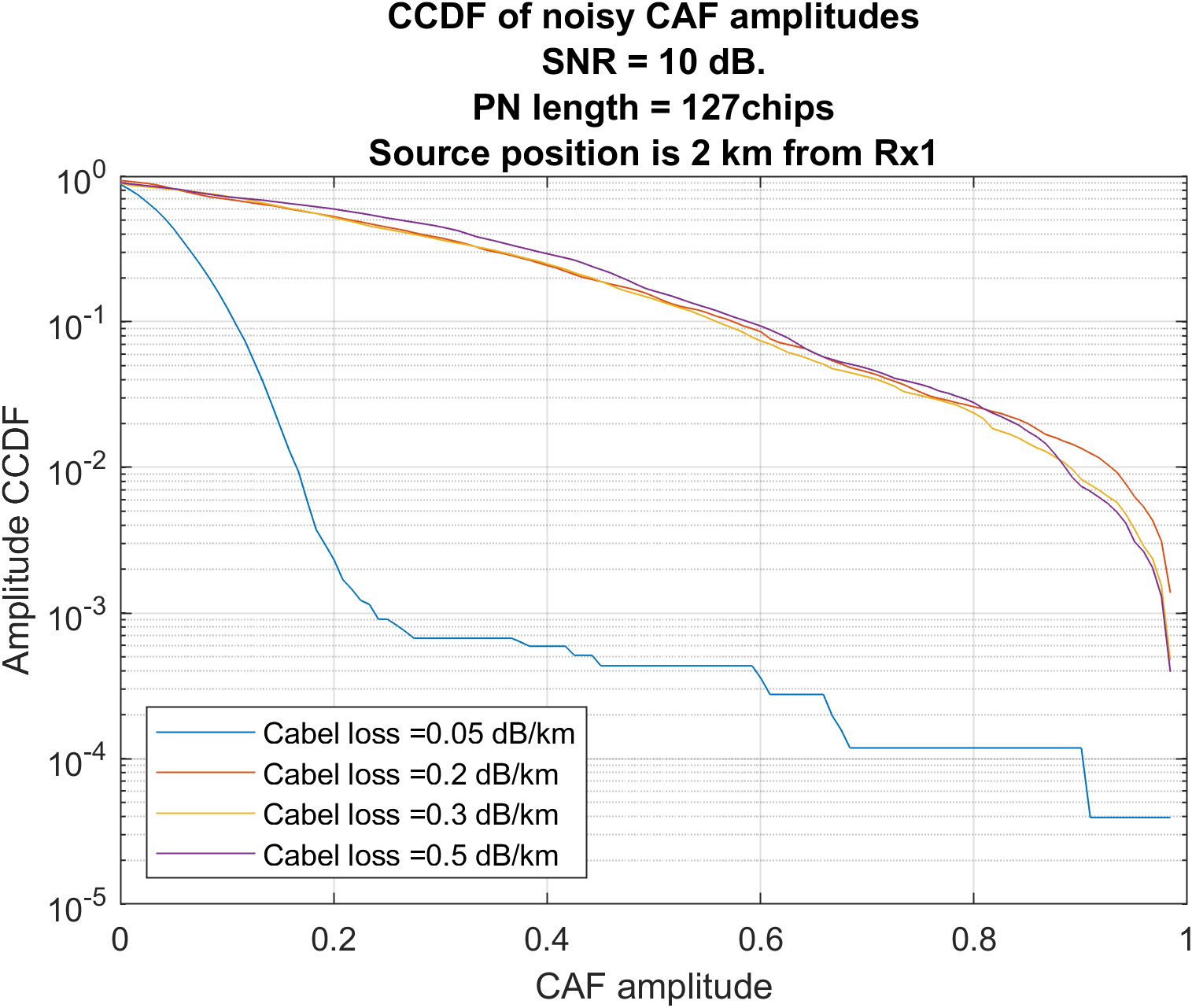}
    \caption{CAF amplitude CCDF at position 2 km for a PN length equal to 127 chips}
    \label{fig:fibL_CCDF_rec1_127}
\end{figure}

The two figures, \cref{fig:fibL_CAF_middle_01} and \cref{fig:fibL_CAF_middle_05}, with the source at the middle of the cable, illustrate the transition from a single peak CAF to a more parasitic peaky CAF as the fiber losses increases. This is due to the reduced SNR of the observed acoustic source at the two optical receivers. 
At the same time the CCDF's for the two different positions,, \cref{fig:fibL_CCDF_middle_127} and \cref{fig:fibL_CCDF_rec1_127}, clearly demonstrate the enhanced sensitivity of the CAF behavior for shift of source position along the fiber.    

\FloatBarrier
\subsection{Varying PN sequence lengths with noise}
 
\begin{figure}[ht!]
    \centering
    \includegraphics[width=0.5\linewidth]{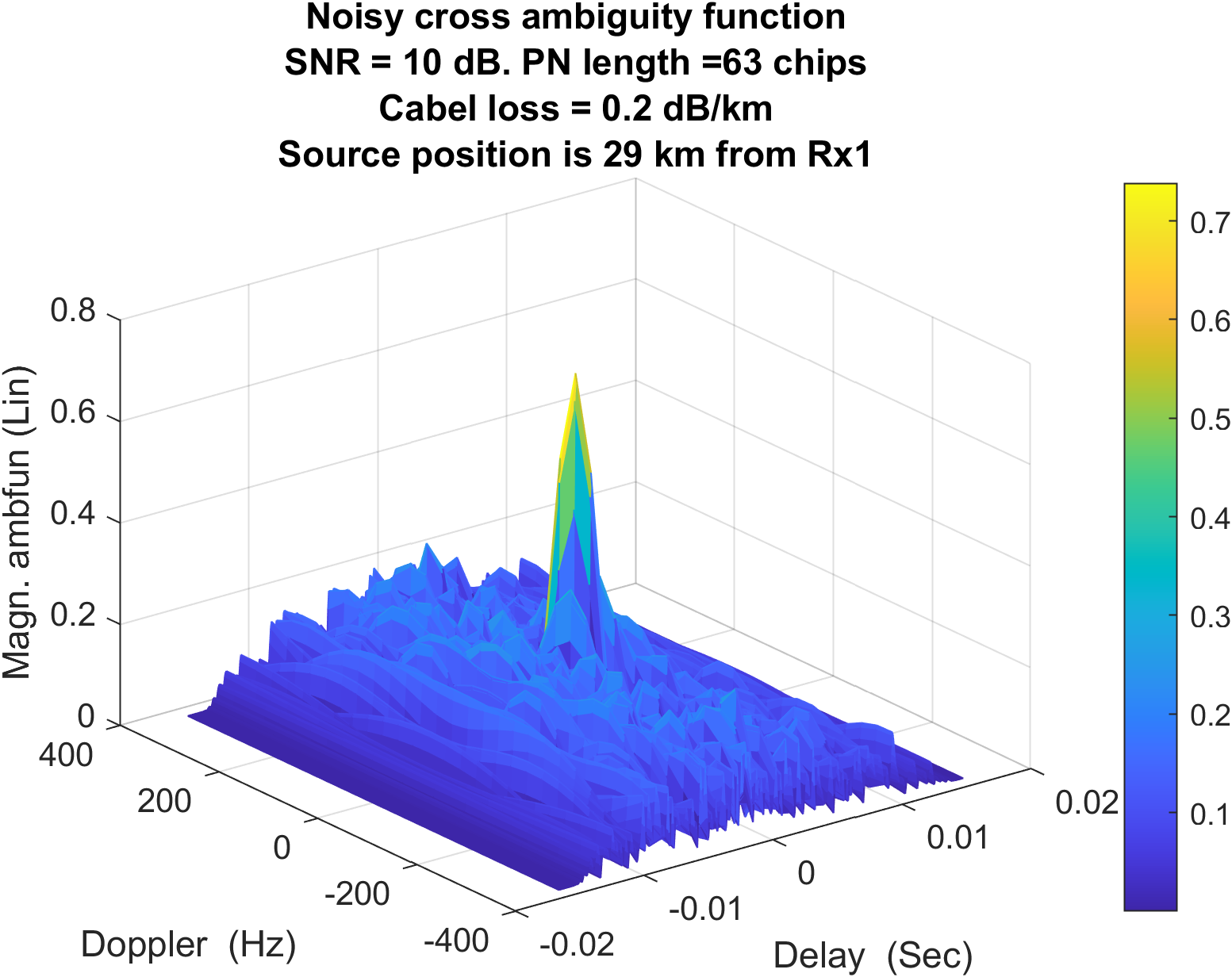}
    \caption{CAF at position 29 km for a PN length equal to 63 chips, and a fiber loss of 0.2 dB/km. .}
    \label{fig:pnL_CAF_63}
\end{figure}

\begin{figure}[ht!]
    \centering
    \includegraphics[width=0.5\linewidth]{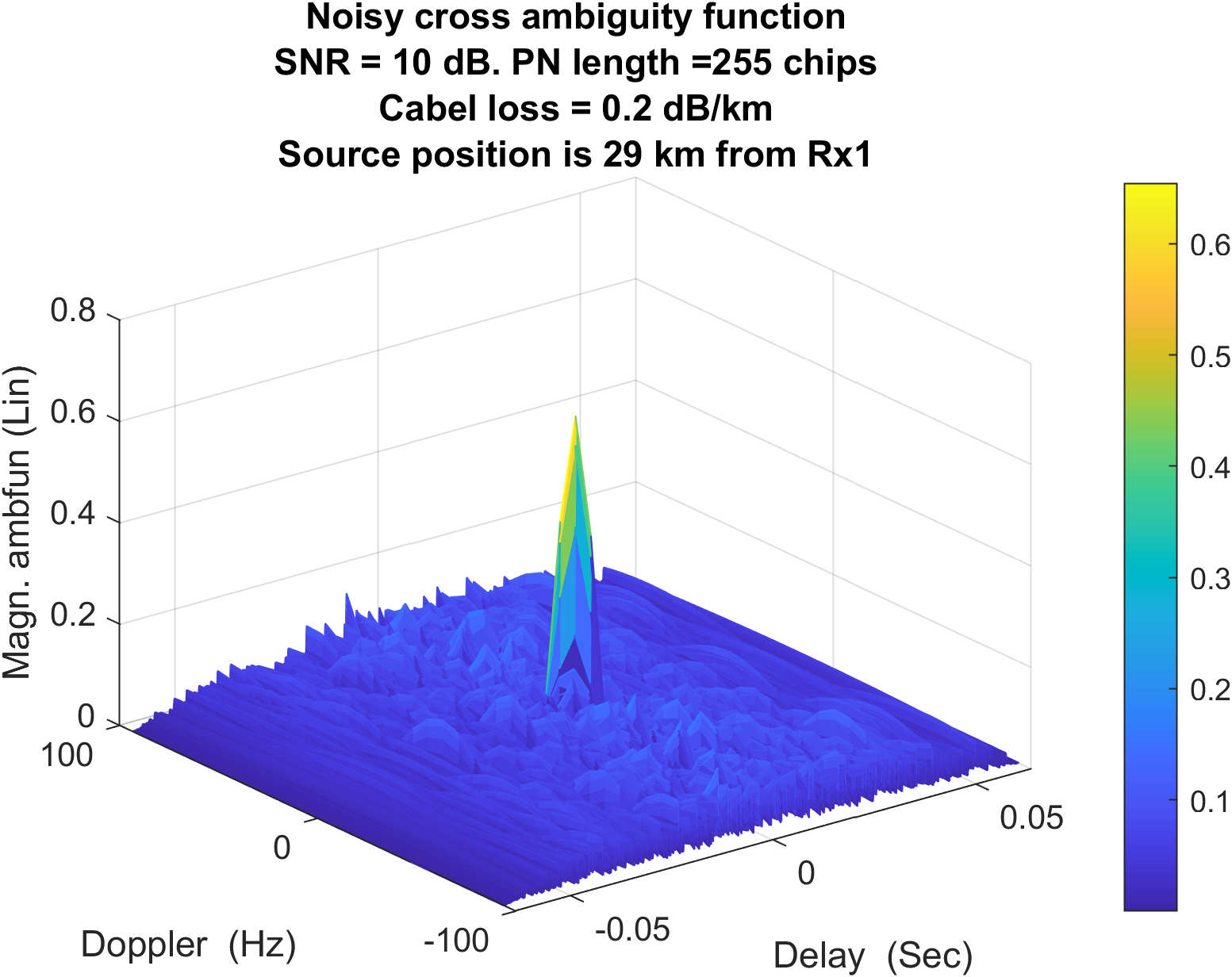}
    \caption{CAF at position 29 km for a PN length equal to 255 chips, and a fiber loss of 0.2 dB/km. .}
    \label{fig:pnL_CAF_255}
\end{figure}

\begin{figure}[ht!]
    \centering
    \includegraphics[width=0.5\linewidth]{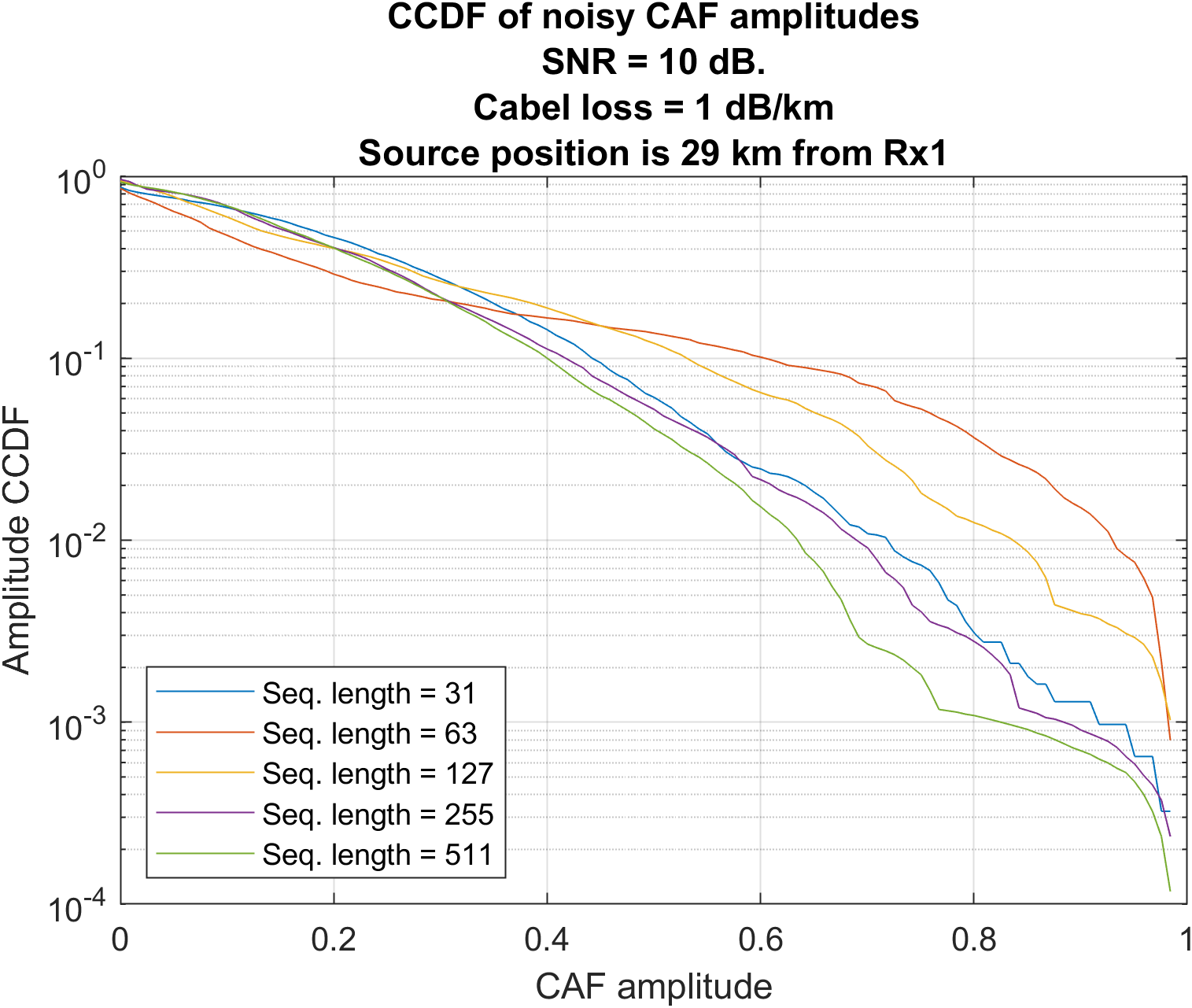}
    \caption{CCDF of CAF at position 29 km for varying PN lengths, and fiber loss of 0.1 dB/km. }
    \label{fig:pnL_CCDF_01_pos29}
\end{figure}

\begin{figure}
    \centering
    \includegraphics[width=0.5\linewidth]{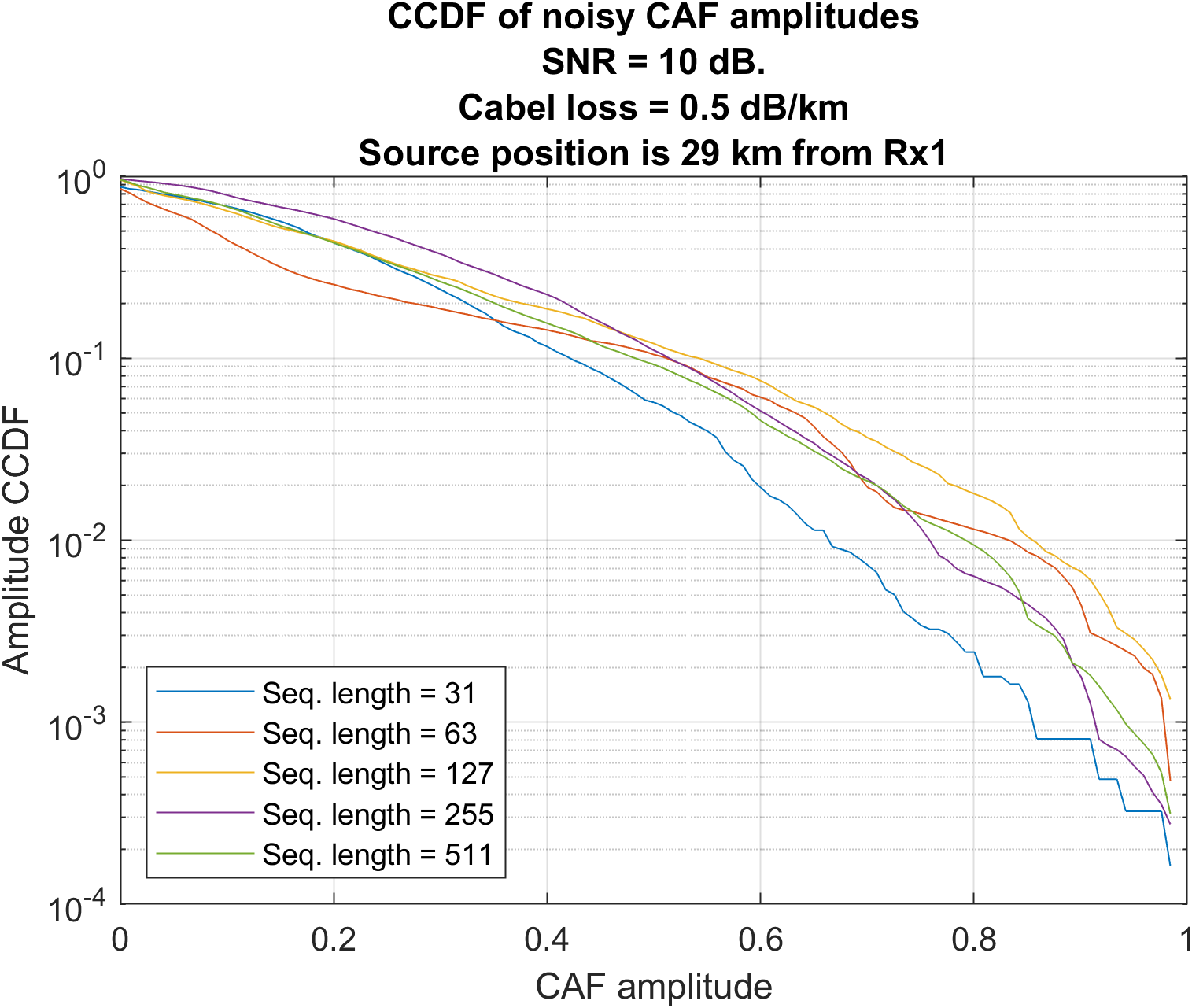}
    \caption{CCDF of CAF at position 29 km for varying PN lengths, and fiber loss of 0.5 dB/km.}
    \label{fig:pnL_CCDF_05_pos29}
\end{figure}

The second important parameter in the CRB formulation, equ. (\ref{equ_delay_crb}), is the number of samples. We have therefore carried out a set of simulations where the length of the PN sequence is varying. The following four figures, \cref{fig:pnL_CAF_63} to \cref{fig:pnL_CCDF_05_pos29}, show the variations of the CAF and CCDF. The source position, 29 km, is at the middle of the cable. The cable loss for the CAF is 0.2 dB/km, while the CCDF is calculated for 0.1 and 0.5 dB/km to underline the influence and importance of the loss factor.      


The influence of the PN length on the CAF, and thereby the ROC hypothesis testing, is clearly seen from \cref{fig:pnL_CAF_63} and  \cref{fig:pnL_CAF_255}, where the parasitic side lobes are lower for 255 chips compared to 63 chips. This also stabilizes the level of the samples of main peak, causing the delay estimate to be unbiased, and to have a lower variance. 

From \cref{fig:pnL_CCDF_01_pos29} and \cref{fig:pnL_CCDF_05_pos29} we can observe differences in the behavior of CCDF's curves. One is due to the loss factors, where a lower loss gives a higher SNR value, which causes the CCDF curves to reflect more the individual properties of the PN sequences them selves, like the variations in single sequence correlation. For larger losses, the noise tends to override the individual properties, and make the curves to converge into a more narrow band. 

The behavior of \cref{fig:pnL_CCDF_01_pos29} shows the importance of selecting both the correct PN length(s), and the respective polynomials.  
\FloatBarrier

\section{Further research and challenges}

The previous sections have demonstrated that it is possible to estimate the position of an acoustic band pass communication signal, being detected and sensed by an optical fiber. However, as also shown, this capability and its spatial resolution, requires that the various system parameters and related technical solutions are bounded within a sub set of possible parameters and solutions, indicated by the CRB bound in equ. (\ref{equ_crb_bound}). E.g. positioning performance improves with increasing acoustic frequencies. In this section we briefly discuss some of these items. It should be noted that each item will have different impact depending upon system configuration, and parameters.  

\begin{itemize}
    \item Fiber acoustic sensing property 
\end{itemize}
 Subsea fiber optical communication cables are  protected to withstand the harsh environment at the ocean bottom. This includes different armament surrounding the optical fiber. In addition to the physical protection, such an armament also causes the acoustic field to be attenuated prior to the interaction with the fiber itself \cite{xiao_review_2022}. The ability of the acoustic wave to penetrate the armament is a function the acoustic carrier frequency, where lower frequencies are less attenuated than higher frequencies. The sensing conversion gain from the acoustic field to the demodulated optical signal amplitude can therefore be expressed as a function of the acoustic carrier frequency, as indicated by the Equ. (\ref{equ_fiber_gain_sens}). 
 
 This frequency dependency of an underwater cable is demonstrated by \cite{potter_distributed_2024}. where the experiments by Potter reveals a fiber response at a few kHz carrier. E.g. \cite{wang_comprehensive_2019} gives an overview of various acoustic optical sensing approaches, some indicating acoustic frequencies higher than 100 kHz. However, the fiber cable constructions are not revealed.

To fully support the acoustic signaling concept proposed in this paper, it is essential to use cables that maintain optimal acoustic coupling across increasing frequencies. This necessitates further research into the performance of commercially available cables, including the potential integration of custom-designed acoustic features or materials to enhance their suitability for this application.  
 
\begin{itemize}
    \item Acoustic wavefront and cable impact
\end{itemize}
Depending upon the distance from the AUV acoustic source to the fiber, and its beam forming, the acoustic signal sensed by the fiber may be shaped as a far field or near field wave front, \cite{liu_near-field_2023}, \cite{elbir_near-field_2025}. In the far field case, the front can be considered to have a near planar, uniform behavior, while a more spherical front represents the near field with a varying signal phase along the fiber. A phase complicating factor is the possibility that the fiber orientation may not be a straight line, but curving, causing additional phase variations along the fiber. This impacts both far - and near field waves.
The acoustic signal from an AUV may be influenced by multipath before reaching the fiber, putting additional requirements on the positioning estimator. A moving AUV will generate Doppler shift, possibly causing a Doppler spread received signal.    

These alternative fields need to be investigated to reveal their relevance and possible level of impact on the system performance.  

An additional effect which may arise and possibly needs attention, represents the imaginary case where the cable is equipped with discrete acoustic resonators forming a long array. In this case the acoustic samples at the two receivers will be two time flipped array output representations of the sensed wave front, causing different array elements time delays for the two receivers.     
   
\begin{itemize}
    \item Acoustic signaling format
\end{itemize}
In wireless communication, orthogonality—or near-orthogonality—is typically achieved across multiple domains, including code, frequency, time, and space. These same domains can be leveraged to maximize the number of independent units accessing a shared optical fiber infrastructure.

To achieve optimal positioning performance across a wide range of system parameters, it is essential that a portion of the acoustic communication frame consists of a long, known code sequence. Pseudo-noise (PN) sequences are particularly suitable for this purpose, as they not only enhance positioning accuracy by an improved detection peak, but also enable Code Division Multiple Access (CDMA).

Furthermore, expanding access to higher acoustic carrier frequencies supports a broader range of applications, including increased data throughput and improved spatial resolution for positioning.

A comprehensive study is required to identify signaling formats that best support the system’s overarching goals—particularly the ability to distinguish and manage multiple units accessing the fiber simultaneously.  
 
\begin{itemize}
    \item CAF peak interpolation  
\end{itemize}
The optimum delay and Doppler estimates are found via interpolating the samples surrounding, and including the maximum peak of the CAF. Alternative approaches may be regular 2D interpolation \cite{jitsumatsu_2d_2023} , or deep learning based if appropriate \cite{rozen_berg_deep_2022}. The goal is to find estimates which approach the CRB for a wide range of SNR values.       

\begin{itemize}
    \item Stable clock references 
\end{itemize}
The delay difference estimation requires a stable clock reference in each of the two receivers. In the case where the receivers are oriented above the sea surface, GNSS clock synchronization is applicable. Otherwise, for sub sea deployed receivers, traditionally crystal oscillators would generally be chosen. However, the new generation of atomic clocks \cite{martinez_chip-scale_2023} , \cite{coccolo_underwater_2022-1}, represent an attractive alternative caused by their enhanced stability performance. 
Choosing clock system solution is an important issue implementing a bistatic positioning system solution, including mutual time synchronization.

\begin{itemize}
    \item Acoustic diversity combining 
\end{itemize}
Given that the two receivers capture distinct versions of the same acoustic signal with uncorrelated additive noise, the signals can be combined using diversity techniques to enhance data detection performance \cite{andrea_goldsmith_wireless_2005}. This principle extends to systems with more than two fibers, where each receiver contributes independently corrupted observations, further improving robustness due to the statistical independence of the noise.

Moreover, the analysis indicates that employing receivers at both ends of the fiber can effectively double the sensing range compared to a single-receiver setup. This diversity gain is achievable when the signal-to-noise ratio (SNR) at each receiver is sufficiently high—typically near the midpoint of the fiber, along its entire length if attenuation is minimal, or in configurations where system parameters are optimized to maintain adequate SNR throughout.  
\begin{itemize}
    \item Colored noise from external sources
\end{itemize}
The capability of optical fibers to detect acoustic communication signals inherently entails the sensing of additional acoustic sources, both natural and anthropogenic \cite{mccarthy_international_2001}. These extraneous sources manifest as colored noise superimposed on the received amplitude signals. In the current context, such noise must be filtered or suppressed to isolate the desired signals. This is achieved by designing the target signals to be orthogonal to the noise sources within a specific domain, thereby enhancing signal fidelity and robustness.  
 
\begin{itemize}
    \item GPS type of data and carrier clock relation
\end{itemize}
We have found that the positioning capabilities improve with the acoustic signaling frequency. The Global Positioning System (GPS) signaling format imposes an integer relation between its carrier frequency and sequence spreading clock rate \cite{wikipedia_note_wikipedia_2025}. This makes it possible to utilize the combined code and carrier phase synchronization for high precision positioning \cite{samper_ppp_2014}. Investigating such an approach for the present two receiver system could be an approach to a enhance the position estimate quality.    

\begin{itemize}
    \item Authentication and security
\end{itemize}
The described system represents one-way acoustic communication. In the case where a communicating unit is a member of a closed set of units, it is mandatory that this closed set can be uniquely identified at the receiver side, both as set and individual member. As a supporting functionality, it is therefore necessary that the acoustic data format supports one-way transmitter authentication, and data security and encryption mechanism. This topic needs to be elaborated for a proper and reliable solution.

\FloatBarrier
\section{Conclusion}
We have presented an approach to estimate the position of an acoustic communication signal sensed by an optical communication fiber, applying the bi-static radar principle, where a dual bi-static solution is formulated using the sensed acoustic signal itself. The topology is a dual fiber cable layout with a separate optical transmitter and receiver at each end. Applying the acoustic signals envelopes, the position is estimated via the observed propagation delay difference between the two receivers. The Cramer Rao bound is deduced illustrating the expected position estimation quality, conditioned on system parameters involving the fiber optical power loss factor. It is found that a larger acoustic bandwidth and higher carrier frequency improve the position resolution along the cable. The position estimate tends to be most accurate near the midpoint of the cable, where the SNR at both receivers is nearly equal. The extent to which performance degrades away from this midpoint is largely determined by the fiber’s optical power loss. Lower cable loss allows for a longer span of cable with good performance compared to cables with higher loss.

The cross ambiguity function (CAF) serves as the maximum likelihood estimator for the delay difference. Simulations are presented to illustrate its statistical behavior, incorporating parameters that influence the Cramér-Rao Bound (CRB) performance. The results show that, for a suitable subset of system parameters, the CAF produces a single dominant peak, enabling reliable position estimation. However, parameters outside this subset may result in parasitic peaks, which can significantly degrade performance

Various challenges or issues need to be elaborated for such an acoustic positioning solution to give proper position quality. Some of these are included and briefly discussed in Section 6.

\section{Acknowledgment}
The authors are pleased to thank senior research scientist Tor Arne Reinen and research manager Bengt Holter at SINTEF Digital for valuable discussions preparing this paper. The work has been carried out with the support of SINTEF Digital.
\FloatBarrier

\bibliography{main_acoustic_position}

\bibliographystyle{ACM-Reference-Format}
\end{document}